\documentclass[twocolumn,merge]{aastex63}
\pdfoutput=1 

\newcommand\Deltaf{\Delta f}  
\newcommand\fhalf{\Delta f_{1/2}} 
\newcommand\Nf{N_\mathrm{f}}
\newcommand\Nt{N_\mathrm{t}}
\newcommand\Son{S^\mathrm{on}}
\newcommand\Soff{S^\mathrm{off}}
\newcommand\tausc{\tau_\mathrm{sc}}    
\newcommand\thetasc{\theta_\mathrm{sc}}    
\newcommand\tscint{t_\mathrm{scint}}   

\shorttitle{Pulsar scintillation spectra}
\shortauthors{Bartel}

\begin{document}

\title{Electron density variations in the interstellar
  medium and the average frequency profile of a scintle from
  pulsar scintillation spectra}

\author{N.\ Bartel}
\affiliation{York University, 4700 Keele St., Toronto, ON M3J 1P3, Canada}

\author{M.~S.\ Burgin}
\affiliation{Lebedev Physical Institute, Astro Space Center, Profsoyuznaya 84/32, Moscow, 117997 Russia}
  
\author{E.~N.\ Fadeev} 
\affiliation{Lebedev Physical Institute, Astro Space Center, Profsoyuznaya 84/32, Moscow, 117997 Russia}

\author{M.~V.\ Popov} 
\affiliation{Lebedev Physical Institute, Astro Space Center, Profsoyuznaya 84/32, Moscow, 117997 Russia}
  
\author{N. Ronaghikhameneh} 
\affiliation{York University, 4700 Keele St., Toronto, ON M3J 1P3, Canada}
\affiliation{University of Alberta, 116 St. \& 85 Ave., Edmonton, AB T6G 2R3, Canada}

\author{T.~V.\ Smirnova}
\affiliation{Lebedev Physical Institute, Pushchino Radio Astronomy Observatory, Pushchino 142290, Moscow region, Russia}

\author{V.~A.\ Soglasnov}
\affiliation{Lebedev Physical Institute, Astro Space Center, Profsoyuznaya 84/32, Moscow, 117997 Russia}

\begin{abstract}
We observed the scintillation pattern of nine
  bright pulsars at 324~MHz and three at 1.68~GHz and
  analyzed the wavenumber spectrum which is related to
  electron density variations of the plasma turbulence of
  the interstellar medium.
  For all pulsars the frequency section of the
  autocorrelation function of the dynamic spectra to at
  least 45\%
  of the maximum corresponds to predictions of scattering
  theories with a range of power-law exponents of the
  wavenumber spectrum of $3.56 \leq \alpha \leq 3.97$ with errors $\leq
  0.05$ and a mean with standard deviation of $3.76\pm0.13$.
  The range includes $\alpha=3.67$ for the Kolmogorov spectrum.
  Similar results although with larger errors were found
  from the Fourier transform of the autocorrelation
  functions down to $\sim 10^{-3}$ of the maximum.  No clear
  case of a distinction between thin-screen and
  extended-medium scattering models was found.  The average
  frequency profile of the scintles can be characterized for
  steep wavenumber spectra with $\alpha\lesssim4$ by a cusp with a
  somewhat rounded peak. For flatter spectra, down to at
  least $\alpha\sim 3.56$ the cusp with its peak becomes more
  pronounced and its decay steepens. We discuss our findings
  in the context of scattering characteristics of the
  interstellar medium.
\end{abstract}


\section{Introduction}  \label{sec:intro} 
Electron density variations in the interstellar medium
  (ISM) scatter radio emission from cosmic
sources.  The variations can be characterized by a
  spatial correlation function. Its Fourier transform is the
  spatial wavenumber spectrum with an exponent, $\alpha$, that
  describes its steepness.  Extensive studies of scattering
effects began with the discovery of pulsars, since pulsars
are quasi point-like sources that provide a coherent flux of
radio waves where the influence of the source structure can
be neglected.  Scattering leads to image blurring, pulse
broadening, and intensity modulation over time and
frequency. The observable parameters characterizing
these phenomena are the scattering angle, $\thetasc$, the
scattering time, $\tausc$, the diffraction scintillation
time, $\tscint$, and the decorrelation bandwidth, $\fhalf $.
Theoretical considerations of scattering effects by, e.g.,
\cite{Lov1970}, \cite{LeeJ1975a, LeeJ1975b, LeeJ1975c},
\cite{rickett1977}, \cite{SSS2003} have established early on
a number of relationships between characteristics of
  electron density variations and observable
  scattering parameters such as those given above.

Previously we reported on a comparison of $\thetasc$ and $
\tausc $ and our estimates of the distances to the effective
scattering screens \citep{gwinn2016, fadeev2018,
  2017MNRAS.465..978P, pop2020}, as well as on the time
characteristics of the scintillation pattern
\citep{Smirnova21}.

In this paper we focus on the frequency characteristics
  of the scintillation pattern.  The frequency section of
the time-averaged autocorrelation functions, $ACF(
  \Deltaf )$, of consecutive pulsar scintillation
  spectra, or dynamic spectra, is often used to derive the
parameter, $ \fhalf $. This section, or its Fourier
transform, $\mathcal{FT}[ACF(\Deltaf)]$, can also be
compared with predictions of thin-screen and extended-medium
scattering models of plasma inhomogeneities in the
intervening ISM, both for a Gaussian and a power-law spatial
wavenumber spectrum with spectral index, $ \alpha $. While
  there have been many observational investigations of the
  parameter, $\alpha$, only very few, and then for only a few
  pulsars, are based on observing {$ACF( \Deltaf )$}
  functions and comparing them to model predictions
  \citep[e. g.,][]{Armstrong1981,Wol1983}.

New insights into the scattering process were obtained with
the development of space VLBI observations of pulsars with
Radioastron which allowed the predicted substructure
\citep{GoodmanN1989} of the visibility function to be
discovered \citep{gwinn2016,2017MNRAS.465..978P}. On long
ground-space baseline projections for which the scattering
disk is fully resolved the visibility function retains
values clearly larger than zero over a range of delays
corresponding to the scattering time, $\tausc$. Since
$\tausc$ is inversely related to $\fhalf$, pulsars with a
broad visibility function are expected to have a narrow
decorrelation bandwidth, $\fhalf$. Therefore, depending on
the time and frequency resolution of the observations, using
the functions $ACF(\Delta f)$ and $\mathcal{FT}[ACF(\Delta f)]$ can be
considered complementary for providing a more complete
analysis of the dynamic spectra.

In this paper we report on an analysis of the dynamic
spectra of 12 pulsars. The time range of the dynamic spectra
is for each pulsar clearly longer than the scintillation
time so that our results refer to an analysis of the
scintillations in the averaging mode \citep{Narayan1992}.
We compare the dynamic spectra with the predictions of thin-screen and extended-medium scattering
models in terms of the wavenumber spectral index, $\alpha$, of the ISM electron density variations, discuss our results with measurements by others, search for hints for an observational preference for either of the models, relate the dynamic spectra to the characteristics of
the substructure in the visibility functions and infer the
average frequency profile of the scintles as a function of $\alpha$.

\section{Observational parameters of the dynamic spectra}  
\label{sec:obs} 
We studied the pulsar dynamic spectra for nine pulsars at a
center frequency of 324~MHz and for three pulsars at
1676~MHz. The data were obtained with VLBI recorders over a
bandwidth of 16~MHz in the context of the space VLBI
Radioastron scientific program \citep{Kardashev2017}.  We
list the pulsars together with their periods, dispersion
measures, galactic coordinates, recording stations, dates of
observations and observing frequencies in
Table~\ref{tab:param0}. The data were already used earlier
in our other studies (refer to "Code" in
Table~\ref{tab:param0}) and more details on these
observations and the first stage of data analysis are given
in our previous papers (see above).

Each dynamic spectrum, $S (f_i, t_j)$ as a function of
frequency, $f$, and time, $t$, consists of $ \Nf \times \Nt $
values, with $1 \leq i \leq \Nf$ and $1 \leq j \leq \Nt$ where $ \Nf $
is the number of frequency channels covering the frequency
range in the bandpass from 316 to 332~MHz or from 1668
  to 1684~MHz, and $ \Nt $ the number of individual spectra
in a given set of observations.  We calibrated the dynamic
spectra as follows:
\begin{equation}        
\label{eq:cal_dyn}
 S (f_i, t_j) = \frac{\Son (f_i, t_j) - \Soff(f_i, t_j)}{ \Soff (f_i, t_j)} 
\end{equation}
and hereafter refer to them simply as $S(f, t)$.
Here, $ \Son $ and $ \Soff $ are the spectra obtained
during the on-pulse and off-pulse windows.  Usually, the
time interval between successive spectra is equal to the
period of the pulsar, but in some cases averaging was
done over several periods to smooth out the intensity
fluctuations from pulse to pulse.  The time span for the
dynamic spectra, as well as $ \Nf $ and $ \Nt $, 
is also listed
in Table~\ref{tab:param0}.

\begin{deluxetable*}{lcrrrrrrccrl}
\tabletypesize{\footnotesize}
\tablecaption{List of pulsars\label{tab:param0}}

\tablewidth{0pt}
\tablehead{
\colhead{PSR} & \colhead{P} & \colhead{DM}
& \colhead{$l_{II}$}
& \colhead{$b_{II}$} & \colhead{Obs.}
& \colhead{$\Nf$} & \colhead{$\Nt$}
& \colhead{Station} &\colhead{Date}
& \colhead{$f$}
& \colhead{Code}
\\
& \colhead{(s)} & \colhead{(pc\,cm${}^{-3}$)} 
&\colhead{(deg)} &\colhead{(deg)}
&\colhead{(min)} &&&& \colhead{(yyyy mm dd)} & \colhead{(MHz)}
\\
\colhead{(1)} & \colhead{(2)} & \colhead{(3)} 
&\colhead{(4)} &\colhead{(5)} &\colhead{(6)}
&\colhead{(7)} &\colhead{(8)}&\colhead{(9)}
&\colhead{(10)}&\colhead{(11)} &\colhead{(12)}
\\
}
\startdata
B0329+54 & 0.714 & 26.7 & 145.0 & -1.2 & 60  & 4096 & 504   & GB & 2012 11 26 & 324 & raes10a \\
B0525+21 & 3.745 & 50.9 & 183.4 & -6.9 & 168 & 512  & 2700  & AR & 2013 09 18 & 1676 &raks02ac \\ 
B0823+26 & 0.531 & 19.4 & 197.0 & 31.7 & 150 & 2048 & 16900 & AR & 2015 03 11 & 324 & rags04aj \\
B0834+06 & 1.274 & 12.8 & 219.7 & 26.3 & 55  & 8192 & 3300  & GB & 2014 12 08 & 324  & rags04ah \\
B0919+06 & 0.430 & 27.3 & 225.4 & 36.4 & 90  & 2048 & 5200  & AR & 2018 05 10 & 324  & rags29p \\
B1133+16 & 1.188 & 4.8  & 241.9 & 69.2 & 120 & 1024 & 6000  & AR & 2018 02 03 & 324  & rags29g \\
B1237+25 & 1.382 & 9.3  & 252.0 & 86.5 & 100 & 512  & 4340  & AR & 2017 12 22 & 324  & rags29c \\
B1642\,--\,03 & 0.387 &35.7  & 14.1  & 26.1 & 90  & 512  & 13200 & WB & 2013 08 09 & 324  & raks02ab\\
B1749\,--\,28 & 0.562 &50.8  & 1.5   & -1.0 & 250 & 192  & 2450  & PA & 2014 05 26 & 1676 & raks02az\\
B1929+10 & 0.226 &3.2   & 47.4  & -3.9 & 100 & 512  & 26000 & AR & 2015 05 05 & 324  &rags04ao\\
B1933+16 & 0.359 &158.5 & 52.4  & -2.1 & 90  & 8192 & 15036 & AR & 2013 08 01 & 1676  &rags02aa\\
B2016+28 & 0.558 &14.1  & 68.0  & -4.0 & 45  & 2048 & 5300  & AR & 2015 05 22 & 324  & rags04aq\\
\enddata
\tablecomments{
Columns are as follows:
(1) pulsar name,
(2) pulsar period,
(3) dispersion measure,
(4) galactic longitude,
(5) galactic latitude,
(6) observing time range for dynamic spectra,
(7) number of frequency channels across the bandpass,
(8) number of spectra over the observing time range,
(9) radio telescope station: AR -- Arecibo, GB -- Green Bank,
    PA -- Parkes, WB -- Westerbork, 
(10) observing date,
(11) center of observing frequency,
(12) code of observing session.
}
\end{deluxetable*}

\begin{figure*}[htb]
\includegraphics{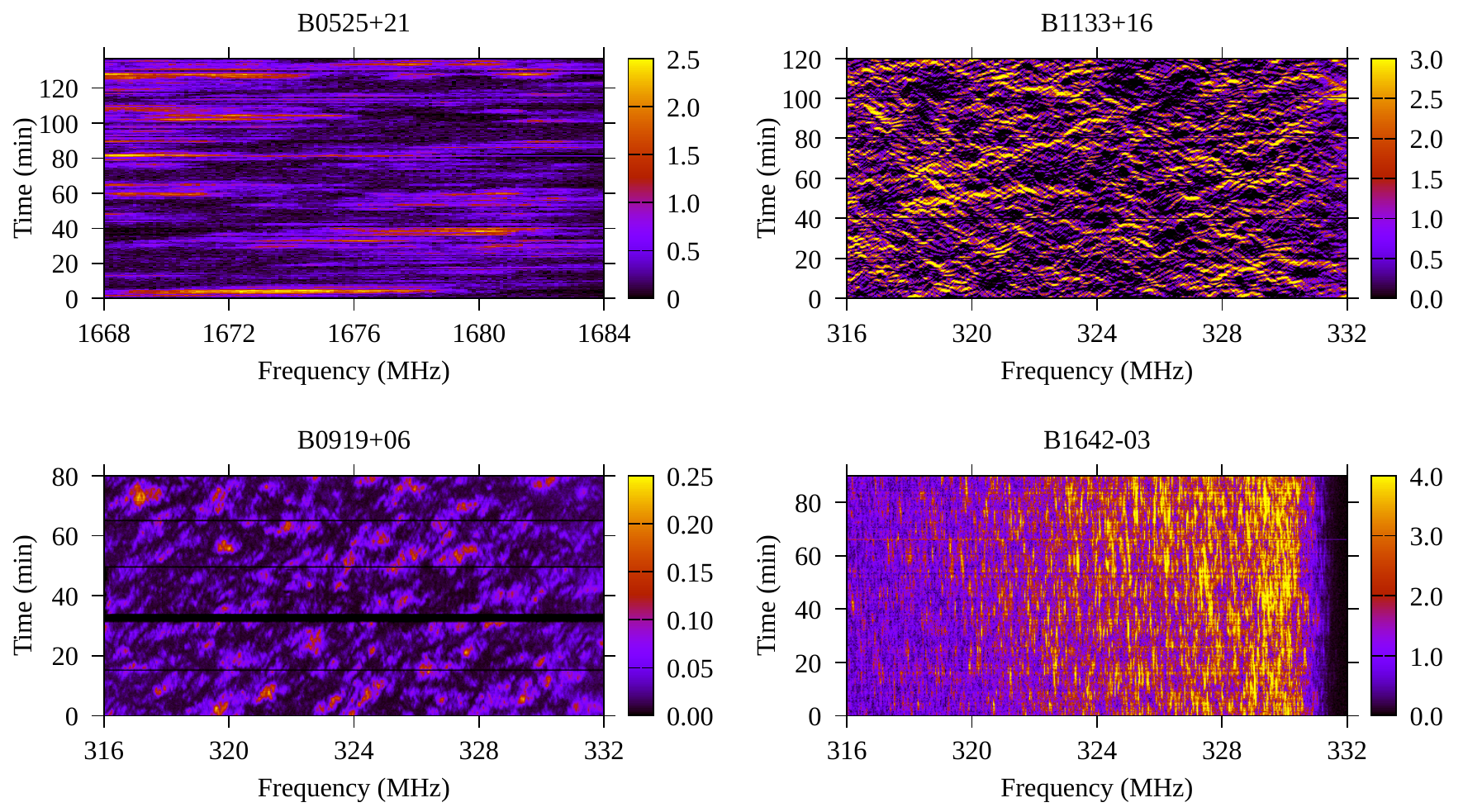}
\caption{Dynamic spectra of four pulsars analyzed in this
  paper. The color-coded intensity as a function of time and
  frequency is given in arbitrary units in the bar on the
  right side of each spectrum.
  \label{fig:4dynspectra}      }
\end{figure*}
In Figure \ref{fig:4dynspectra} we give examples of
  dynamic spectra for four pulsars: B0525+21, B0919+06,
  B1133+16, and B1642-03. The dynamic spectra for the other
  eight pulsars in our sample were already published earlier
  \citep{popov2016, popov2021, fadeev2018}.  One can see
islands of increased intensity distributed randomly over the
frequency-time domain. These are the peaks of diffractive
structure, which we refer to as 'scintles.' In particular we
are interested in the decorrelation bandwidth, $\fhalf$, the
frequency section of the autocorrelation function, $ACF(\Delta
f)$, as well as the Fourier transform of this function,
  their relation to parameters of the turbulent
interstellar plasma and the average frequency profile of the
scintles.

\section{Analysis of the dynamic spectra}
\subsection{Frequency structure of the dynamic spectra and their relation to the turbulent plasma of the intervening ISM}  \label{subsec:rel-betw}

Scintillations of pulsars are related to electron density
fluctuations in the ISM which can be characterized by a
spatial correlation function. Its Fourier transform is the
spatial wavenumber spectrum, $P(q)$. If the magnitude of the
three dimensional wavenumber, $q$, is within a limited range
of wavenumbers with inner and outer boundaries, then $P(q)$
can be described as a power-law, $P_{\mathrm P}(q)=C_{n}q^{-\alpha}$,
with $\alpha < 4$ \citep{Roman1986} including $\alpha = 11/3$ for a
Kolmogorov spectrum.  For a Gaussian the spectrum is,
$P_{\mathrm G}(q)=G_{n}\exp(-(q/q_{0})^{-2})$ corresponding to $\alpha =
4$.  The value of $\alpha$ can be obtained from the 
two-dimensional autocorrelation function of $S(f,t)$ at zero
time lag, $ACF(\Deltaf, \Delta t=0)$. For simplicity we refer to
this frequency section after normalization as $ACF(\Deltaf)$
with
\begin{equation}
ACF(\Deltaf)=
\left\langle (S(f,t)) (S(f+\Deltaf,t)) \right\rangle_t.
\end{equation}
The frequency section, $ACF(\Deltaf)$, is predicted to
depend on $\alpha$. For a thin-screen model of the ISM analytical
solutions exist for the Gaussian and power-law models that
can be fit to the autocorrelation functions. For an extended-medium model of the ISM \citep[e.g.][]{LeeJ1975a} analytical
solutions for the frequency section of the ACF exist for the
Gaussian spectrum but only numerical solutions for the
power-law spectrum \citep{LeeJ1975b}.

Instead of computing the frequency section, $ACF(\Deltaf)$,
for comparison with predictions, some authors preferred to
compare the Fourier transform of this ACF with
predictions. Both approaches can be considered
complementary. In this paper we use both approaches by
following \citep{Wol1983} concerning the computation of the
predictions of the extended-medium scattering model and
\citet{Armstrong1981} concerning the thin-screen model.

\subsection{Determination of the power-law spectral index, \texorpdfstring{$\alpha$}{alpha}, of the wavenumber spectrum}
\label{subsec:freq_section}

We studied 12 pulsars by analyzing the frequency section,
$ACF(\Deltaf)$, of the dynamic spectra.  From $ACF(\Deltaf)$
we determined the decorrelation bandwidth, $\fhalf$, as the
half-width at half maximum (HWHM) and list the values for each
pulsar in Table~\ref{tbl:results}.  The fractional
statistical (rms) error can be estimated as the inverse
square root of the number of scintles in each dynamic
spectrum \citep{backer1975}. An approximate expression is
given by $\sigma_\mathrm{stat} = \left(f_s \frac{BT_\mathrm{obs}}{\fhalf
  \tscint} \right)^{-1/2}$ where $ f_s$ is the filling
factor, $B$ the receiver bandwidth, $T_\mathrm{obs}$ the observing
time, and $t_{scint}$ the scintillation time. Assuming
$f=0.5$ and taking values for $t_\mathrm{scint}$ from
\citet{Smirnova21} we list $\sigma_\mathrm{stat}$ in
Table~\ref{tbl:results}. For a consideration of possible
systematic errors, see \citet{Smirnova21}.

\begin{deluxetable*}{lrccccccc}
\tablecaption{Results of data reduction\label{tbl:results}}
\tablewidth{0pt}
\tablehead{
\colhead{PSR} & \colhead{$\fhalf$}
&\colhead{$\Delta \tau_{1/2}$}
&\colhead{$\sigma_\mathrm{stat}$}
&\colhead{$\alpha_{ACF}$}
&\colhead{$rms_{ACF}$}
&\colhead{$\alpha_{\mathcal{FT}(ACF)}$}
& \colhead{$rms_\mathrm{scat.\,type}$}
& \colhead{Scat. type}
\\
& \colhead{(kHz)} &$(\mu s)$ &\colhead{$(10^{-2})$} &&\colhead{$(10^{-2})$}& &\colhead{$(10^{-2})$}
\\
\colhead{(1)} & \colhead{(2)} & \colhead{(3)} 
&\colhead{(4)} &\colhead{(5)}  &\colhead{(6)} &\colhead{(7)} &\colhead{(8)} &\colhead{(9)} 
}
\startdata
B0329$+$54 & 30  & 7.90 & 1.0 & 3.67$\pm$0.02  & 0.05 & 3.80$\pm$0.05  & 0.15 &  K \\
B0525$+$21 & 4000 & 0.04 & 8.6 & 3.73$\pm$0.02  & 0.5  & 3.80$\pm$0.05 & 3.1/4.2/4.2 &  K/E/L\\
B0823$+$26 & 180  & 0.83 & 1.2 & 3.56$\pm$0.02  & 0.2  & 3.60$\pm$0.05  & 1.8 &  K \\
B0834$+$06 & 400  & 0.55 & 5.8 & 3.74$\pm$0.02  & 0.2  & 4.00$\pm$0.10  & 1.8 &  K \\
B0919$+$06 & 170  & 1.15 & 1.9 & 3.59$\pm$0.02 & 0.2  & 3.70$\pm$0.10  & 1.7 &  K \\
B1133$+$16 & 96   & 2.50 & 0.7 & 3.80$\pm$0.02  & 0.3  & 4.00$\pm$0.10  & 2.3/3.7 &  K/L \\
B1237$+$25 & 2300 & 0.11& 3.3 & 3.90$\pm$0.04 &0.7&3.80$\pm$0.10  & 2.7/3.8 &  G/E \\
B1642$-$03 & 930  & 0.20 & 4.0 & 3.97$\pm$0.05  & 1.4  & 3.80$\pm$0.20  & 3.0/3.7  & G/E\\
B1749$-$28 & 690  & 0.31 & 3.5 & 3.78$\pm$0.02 & 0.2  & 3.90$\pm$0.10  & 3.7/3.8/3.9 & E/L/K \\
B1929$+$10 & 970  & 0.39 & 6.8 & 3.85$\pm$0.02 & 0.4  & 3.90$\pm$0.10  & 2.8 &  E \\
B1933$+$16 & 120  & 6.85 & 1.1 & 3.64$\pm$0.04 & 0.7  & 3.80$\pm$0.10  & 1.7 & K \\
B2016$+$28 & 71   & 2.04 & 4.9 & 3.92$\pm$0.02 & 0.2  & 3.90$\pm$0.05  & 1.1/1.7 &  L/E \\
\enddata 
\tablecomments{
Columns are as follows:
(1)~pulsar name; 
(2)~HWHM of the ACFs of the 
    dynamic spectra at zero time lag (the smaller value for for B1643-03 of $\sim$ 6 kHz is not considered in this paper); 
(3)~HWHM of the FT(ACF)s;
(4)~fractional rms uncertainty of values in columns (2) and (3);
(5)~parameter, $\alpha$, from the fit of the extended-medium scattering        model to the observed ACFs down to 0.45 of the maximum. The errors         were calculated from the statistical errors derived from column (6)        added in quadrature with systematic errors estimated from slight           deviations of the ACF from the model (same values for extended-medium      and thin screen models);
(6) the rms values in units of correlation coefficients for the ACFs with      respect to the fit model;
(7) ~parameter, $\alpha$, from the fit of the extended-medium
    scattering model to the Fourier transforms of the observed ACFs for the delay range from 0.2 to 20 $\Delta\tau/\Delta \tau_{1/2}$ or down to $\sim 10^{-3}$ of the maximum (same values for extended-medium and thin screen models). The errors are largely systematic errors estimated from non-random deviations of $\mathcal{FT}(ACF)$ from the models;
(5, 7) values including errors larger than 4 are still valid but nonphysical within the context of the models and may indicate deviations from them;    
(8) the smallest rms values in units of $(10^{-2})$ of the maximum of unity of the observed ACFs with respect to the analytical $ACF(f)$ functions in column (9) from Table~\ref{tbl:bells} up to $2\times \Delta f/\Delta f_{1/2}$ (see Figure~\ref{fig:fig7}); 
(9) the best matching functions from the four functions from Table ~\ref{tbl:bells}, $ACF(G)$ to $ACF(K)$.
}
\end{deluxetable*}

\begin{figure*}[htb]
\includegraphics{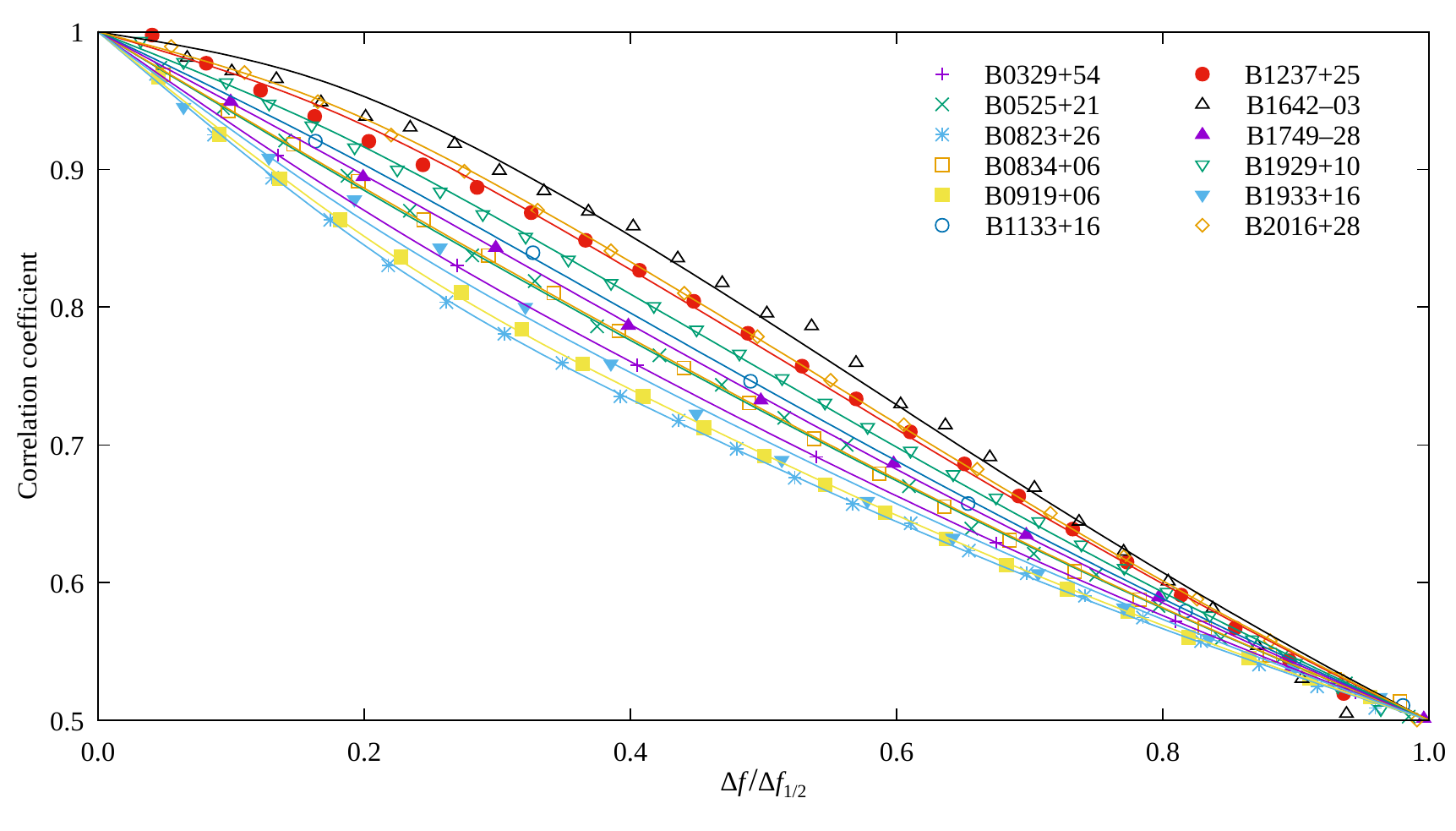}  
\caption{Comparison of the predictions from the
  extended-medium model of \cite{LeeJ1975b} with computed
  autocorrelation functions for our 12 pulsars.
 The
  predictions are plotted as solid lines for indices of $\alpha$
  corresponding to the best-fitting models.
  The ACF's are
  plotted as data points with different symbols for each of
  the pulsars listed on the upper right.
\label{fig:fig2}
}
\end{figure*}

We fit the functions with a curve corresponding to the
predictions for the extended-medium model. 
The differences
between the predictions for $ACF(\Delta f)$ for the thin-screen
and the extended-medium models are small for the inner part
but larger for the tail of the functions. Using, for
instance, the solutions for the Gaussian spectrum ($\alpha=4.0$)
\citep{Lov1970, ChasheiS1976, Lerche1979} with a maximum at
unity and scaling them so that the values for $\Delta f_{1/2}$ are identical, the
functions differ by less than 0.002 down to the HWHM and by
less than 0.007 down to 20\% of the maximum. 
These differences are not large enough to be considered for
our analysis of the inner part of $ACF(\Delta f/\Deltaf_{1/2})$
but have to be considered for the tail of the ACFs where
they could perhaps become measurable. Although for some
pulsars a good fit could be obtained even for the tail of
$ACF(\Deltaf/\Deltaf_{1/2})$, a consistently good fit for
all pulsars was only obtained up to frequency lags,
$\Deltaf/\Deltaf_{1/2} \sim 1.2$, where
$ACF(\Deltaf/\Deltaf_{1/2})=0.45$. For this range, the
fitted values of $\alpha$ for the two scattering models were the
same within our sensitivity limits.  We list the values of
$\alpha$ together with their uncertainties in Table~\ref{tbl:results}.

For a summary view of our 12 measured $ACF(\Delta
f/\Deltaf_{1/2})$ functions we plot their inner portions
down to half of the maximum and
compare them with the fit model predictions in
Figure~\ref{fig:fig2}. It can be seen that the value of $\alpha$
is mostly determined by the shape of the inner part of
$ACF(\Delta f/\Deltaf_{1/2})$. All the functions are within a
range of fit parameters, $\alpha$, between 3.56 for the lower
curve of PSR B0823+26 and 3.97 for the upper curve of PSR
B1642-03.  The mean with standard deviation is $\langle\alpha_{ACF}\rangle=3.76\pm0.13$ and is therefore close to $\alpha_K= 3.67$ of the
Kolmogorov spectrum.  The quality of the fit varies. We list
rms variations of the observed ACFs from the model for
each pulsar in Table~\ref{tbl:results}. The fits are for
most pulsars quite good with rms values smaller or equal to
0.005 on the scale of the ACF functions down to 0.45 of the
maximum of unity. Larger deviations with rms values of 0.007
to 0.014 were found for PSRs B1237+25, B1642-03 and
B1933+16. These are mostly not random variations but to a
large part systematic deviations from the model.

\begin{figure*}[htb]
\includegraphics{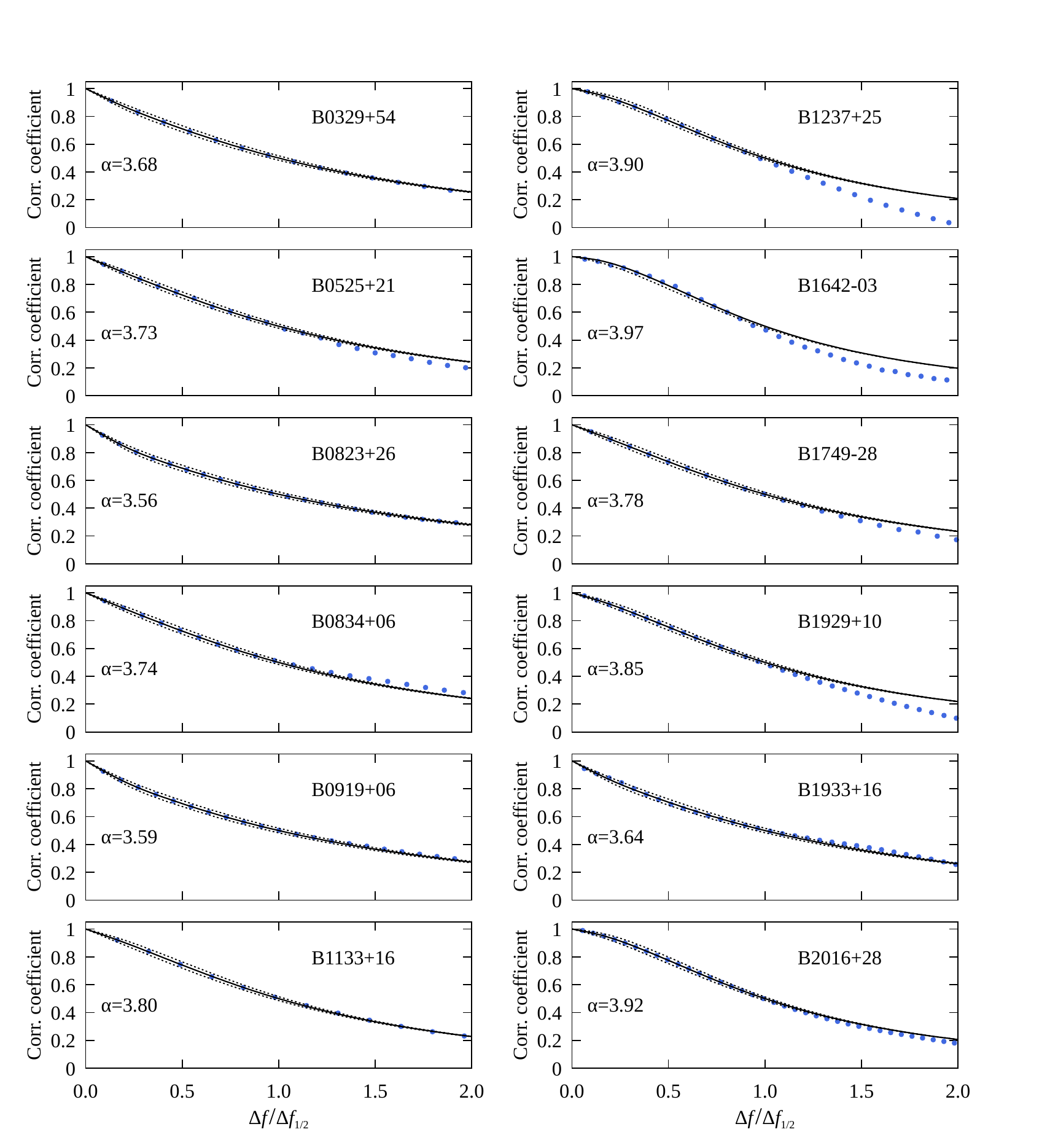}
\caption{Comparison of the $ACF(\Delta f)$ functions derived from
  the observations (blue dots) with the theoretical
  predictions for the extended-medium scattering theories
  with power-law spectral index, $\alpha$, as a fitting
  parameter. The solid line shows the best-fitting model for
  the inner part of the $ACF(\Delta f)$ down to 45\% consistently
  for all pulsars. The surrounding dotted lines indicate
  $\Delta\alpha=\pm 0.05$ deviations from the best-fit value for
  illustration purposes. For better comparison all functions
  were normalized by their HWHM frequency lags, $\Delta f_{1/2}$.
\label{fig:fig3}
}
\end{figure*}

To inspect the fits in more detail down to lower correlation
coefficients, we plot $ACF(\Delta f/\Deltaf_{1/2})$ with the
models from Figure~\ref{fig:fig2} for each of the 12 pulsars
separately for frequency lags up to 2 $\times$ $\Delta f_{1/2}$ in
Figure~\ref{fig:fig3}. For about half the pulsars the data
are still well fit by the model over the wider range of
frequency lags. However significant deviations are now
apparent for the other half.

Some of these deviations in the tail of $ACF(\Delta
f/\Deltaf_{1/2})$ could be due to technical effects such as
1) superposition of the contribution from neighboring
scintles; 2) limited band of the receiver filter which
distorts or cuts off wide diffraction structures; 3)
inaccuracy of the determination of the off-pulse level for
the computation of ACFs and 4) relatively low
signal-to-noise ratios (SNRs).  Some of these effects if not all could possibly increase the ACFs toward large lags. Such increase
can be seen for PSRs B0834+06 and B1133+16. For the other
pulsars with deviations, the tail of the ACFs is lower than
the model predictions. In these cases the deviations are
likely indications of deficiencies of the scattering models.

A complementary method to estimate $\alpha$ values is to fit the
models to the Fourier transform of the ACFs,
$\mathcal{FT}(ACF)$ \citep[e.g.,][]{Armstrong1981, Wol1983}.
We computed $\mathcal{FT}[ACF(\Deltaf)]$ by using the full
range of frequency delays of $\pm4$ MHz. To suppress
fluctuations, we applied to the observed ACF$(\Deltaf)$
functions a Gaussian filter with a half-width at the 1/e
level of $4\times\Deltaf_{1/2}$. Only for
PSR B0525+21 with its relatively large value of
$\Deltaf_{1/2}$ we used a four times narrower Gaussian.

\begin{figure*}[htb]
\centering
\includegraphics{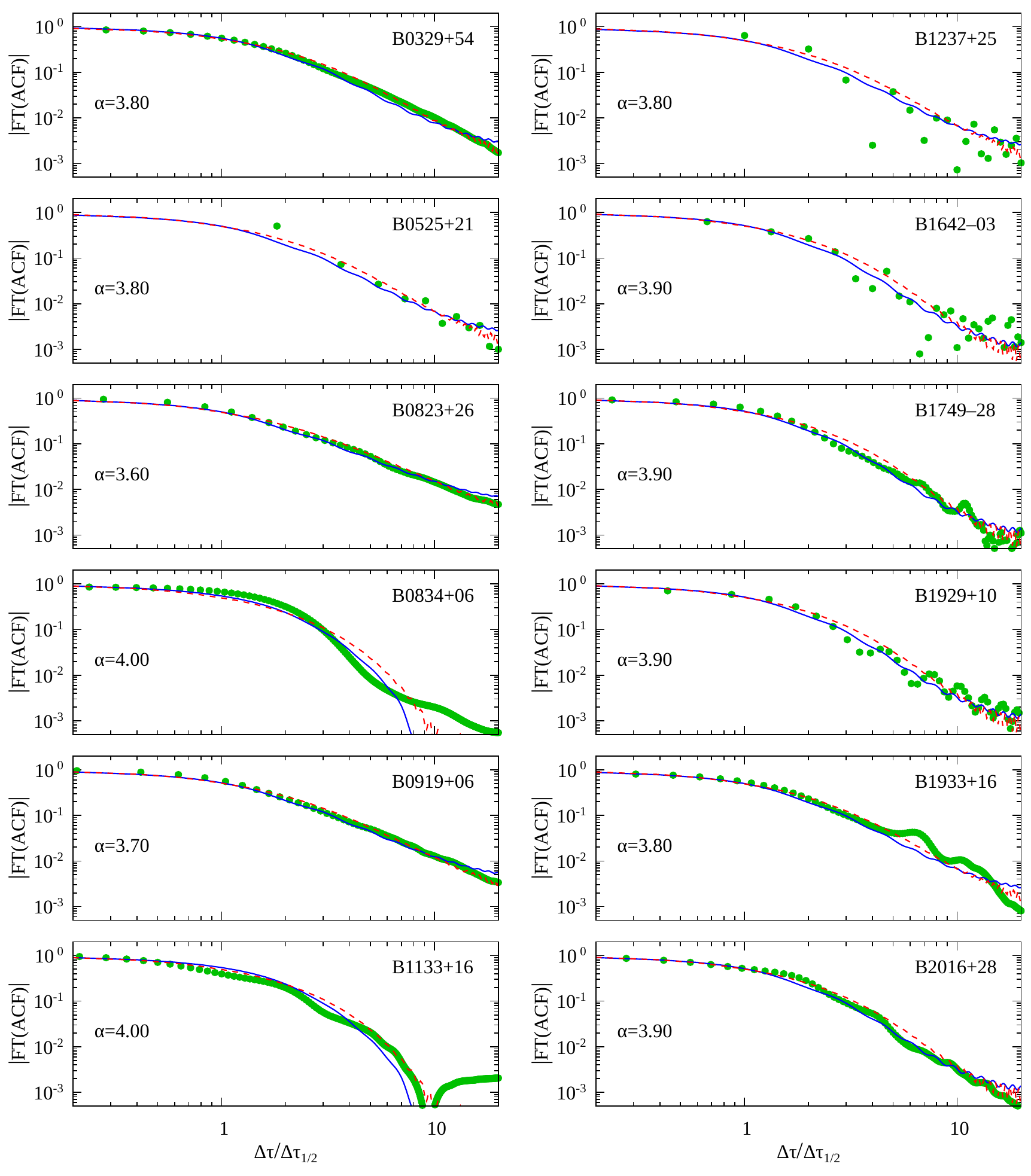}
\caption{The magnitude of the Fourier transform of the observed ACFs, \textbar
  FT(ACF)\textbar, (green
  dots), fit by the theoretical predictions for the
  extended-medium scattering theory with rounded power-law spectral
  index $\alpha$ (solid blue lines) and for the thin-screen model
  (dashed red lines) with the same value for $\alpha$. Scattered
  points in the tail of some functions are due to
  noise. Delay is given as $\Delta \tau$ with $\Delta \tau_{1/2}$ as the
  HWHM listed in Table ~\ref{tbl:results}.
 \label{fig:fig4}}
 \end{figure*}

A comparison of the observed functions
$\mathcal{FT}[ACF(\Deltaf)]$ of all pulsars with the
theoretical predictions for the extended-medium and for the
thin-screen models is shown in Figure~\ref{fig:fig4}. In
this subsection we focus on the fit of the former to the
FT(ACF)s.  All FT(ACF) functions are normalized in amplitude
and in width. Their HWHM values, $\Delta\tau_{1/2}$, are listed in
Table~\ref{tbl:results}.

The values of $\alpha_{\mathcal{FT}(ACF)}$ and their estimated
errors are listed in Table~\ref{tbl:results} (column
(7)). The mean with standard deviation is
$\langle\alpha_{\mathcal{FT}(ACF)}\rangle=3.83\pm0.12$, larger by 0.07 than
$\langle\alpha_{ACF}\rangle$. The mean of the magnitudes of the differences to those obtained
from the fits to the ACFs with standard deviation is
$\langle|\alpha_{ACF}-\alpha_{\mathcal{FT}(ACF)}|\rangle=0.11\pm0.07$. In general
the values are fairly consistent within the errors given
that the fitting ranges are quite different. The largest
discrepancy is 0.26 for PSR B0834+06. The difference could
perhaps at least partly be due to the large deviations from
the model in the ACF for lags outside the ACF fitting range.

\subsection{Comparison with other estimates of
  \texorpdfstring{$\alpha$}{alpha}}
The exponent, $\alpha$, of the wavenumber power-law spectrum of electron density
variations can be estimated by several methods. In
Table~\ref{tbl:compare} we compare our estimates from
Table~\ref{tbl:results} in columns (5) and (7) with those
from other authors derived from the Fourier transform of
the frequency section of the ACF's, structure function,
and frequency dependence of the decorrelation bandwidth.

The most similar analysis to ours was done by
\citet{Wol1983} (column (4)) and \citet{Armstrong1981}
(column (5)) with data taken some 30 years before ours who
fit an extended-medium and a thin-screen scattering model,
respectively, to the FTs of the ACFs. Compared with the
values of \citet{Wol1983}, our corresponding values in
column (3) are all consistent with theirs within our errors.
Compared with the values of \citet{Armstrong1981} which are
given with relatively large ranges, ours are within their
ranges or slightly above but exclude the range below
$\alpha=3.5$.

The values in the other columns were obtained with different
analyses schemes. The most interesting for a comparison are
those listed in column (6) which are based on the same data
as ours in columns (2) and (3). They are derived from the
structure function of the time section of the ACFs and are
therefore best compared with our values in column (2).  For
five pulsars, the differences in $\alpha$ are $\leq$0.10. For the
other four pulsars, ignoring B2016+28 (see caption of Table
\ref{tbl:compare}), the differences are as large as 0.51 as
for B1237+25.

The last column lists estimates of $\alpha$ obtained from a fit
of the decorrelation bandwidth as a function of observing
frequency. They are consistent with our values in column (2)
within 1.5 times their larger uncertainties.  We comment on
these comparisons in Section~\ref{sec:disc}.

\begin{deluxetable*}{lcccccc}
\tablecaption{Comparison of estimates of $\alpha$\label{tbl:compare}}
\tablewidth{0pt}
\tablehead{
     \colhead{PSR}   
    &\colhead{$\alpha_{ACF(\Deltaf)}$}
    &\colhead{$\alpha_{\mathcal{FT}(ACF)}$}
    &\colhead{$\alpha_{\mathcal{FT}(ACF)}$}
    &\colhead{$\alpha_{\mathcal{FT}(ACF)}$}
    &\colhead{$\alpha_\mathrm{SF}$}
    &\colhead{$\alpha_{\Delta f_{1/2} (f)}$}
    \\
     \colhead{(1)} & \colhead{(2)} & \colhead{(3)} &
    \colhead{(4)} & \colhead{(5)} &\colhead{(6)} &\colhead{(7)} 
}
\startdata
B0329$+$54 & $3.68\pm0.02$ & $3.80\pm0.05$ & 3.85 & 3.7-4.0   & $3.67\pm0.01$ & $3.41_{-0.25}^{+0.41}$   \\
B0525$+$21 & $3.73\pm0.02$ & $3.80\pm0.05$ &      &           &                   \\
B0823$+$26 & $3.56\pm0.02$ & $3.60\pm0.05$ &      &           & $3.66\pm0.01$      \\
B0834$+$06 & $3.74\pm0.02$ & $4.00\pm0.10$ &      &           & $3.528\pm0.006$   \\
B0919$+$06 & $3.59\pm0.02$ & $3.70\pm0.10$ &      &           & $3.57\pm0.01$     \\
B1133$+$16 & $3.80\pm0.02$ & $4.00\pm0.10$ & 3.90 & 3.4-3.8   & $3.86\pm0.01$     \\
B1237$+$25 & $3.90\pm0.04$ & $3.80\pm0.10$ &      &           & $3.39\pm0.01$     \\
B1642$-$03 & $3.97\pm0.05$ & $3.80\pm0.20$ &      & 3.2-3.9   &               &$3.63_{-0.21}^{+0.29}$    \\
B1749$-$28 & $3.78\pm0.02$ & $3.90\pm0.10$ &      &           & $3.82\pm0.01$  & $3.50_{-0.14}^{+0.18}$  \\
B1929$+$10 & $3.85\pm0.02$ & $3.90\pm0.10$&      &           & $3.65\pm0.02$    \\
B1933$+$16 & $3.64\pm0.04$ & $3.80\pm0.10$ & 3.90 &           & $3.18\pm0.01$  & $3.80_{-0.12}^{+0.14}$  \\
B2016$+$28 & $3.92\pm0.02$ & $3.90\pm0.05$ &      &           & $\;\,3.36\pm0.02^*$    \\
\enddata 
\tablecomments{
Columns are as follows:
(1)~pulsar name, 
(2)~taken from Table~\ref{tbl:results}, column (5);
(3)~taken from Table~\ref{tbl:results}, column (7);
(4)~estimates of $\alpha$ of the extended medium power-law
    model from the Fourier transform of the frequency ACFs,
    PSRs B0329+54 (327 and 480 MHz), B1133+16 (327 MHz),
    B1933 (1416 MHz) \citep{Wol1983};
(5)~estimate of $\alpha$ of the thin-screen power-law
    model from the Fourier transform of the frequency ACFs,
    PSRs B0329+54 (340, 408 MHz), B1133+16 (340 MHz),
    B1642-03 (340 MHz)  \citep{Armstrong1981};
(6)~estimate of $\alpha$ of the power-law
    spectrum from the structure function of the time section
    of the ACFs, with '$^*$' indicating a measurement biased
    toward lower values \citep{Smirnova21};
(7)~estimate of $\alpha$ of the power-law model from the
    decorrelation bandwidth, $\Deltaf_{1/2}$, as a function
    of observing frequency in the range 80-8100 MHz,
    \citep{cwb1985}.
}
\end{deluxetable*}

\subsection{Thin screen or extended medium?}

\begin{figure*}[htb]
\includegraphics{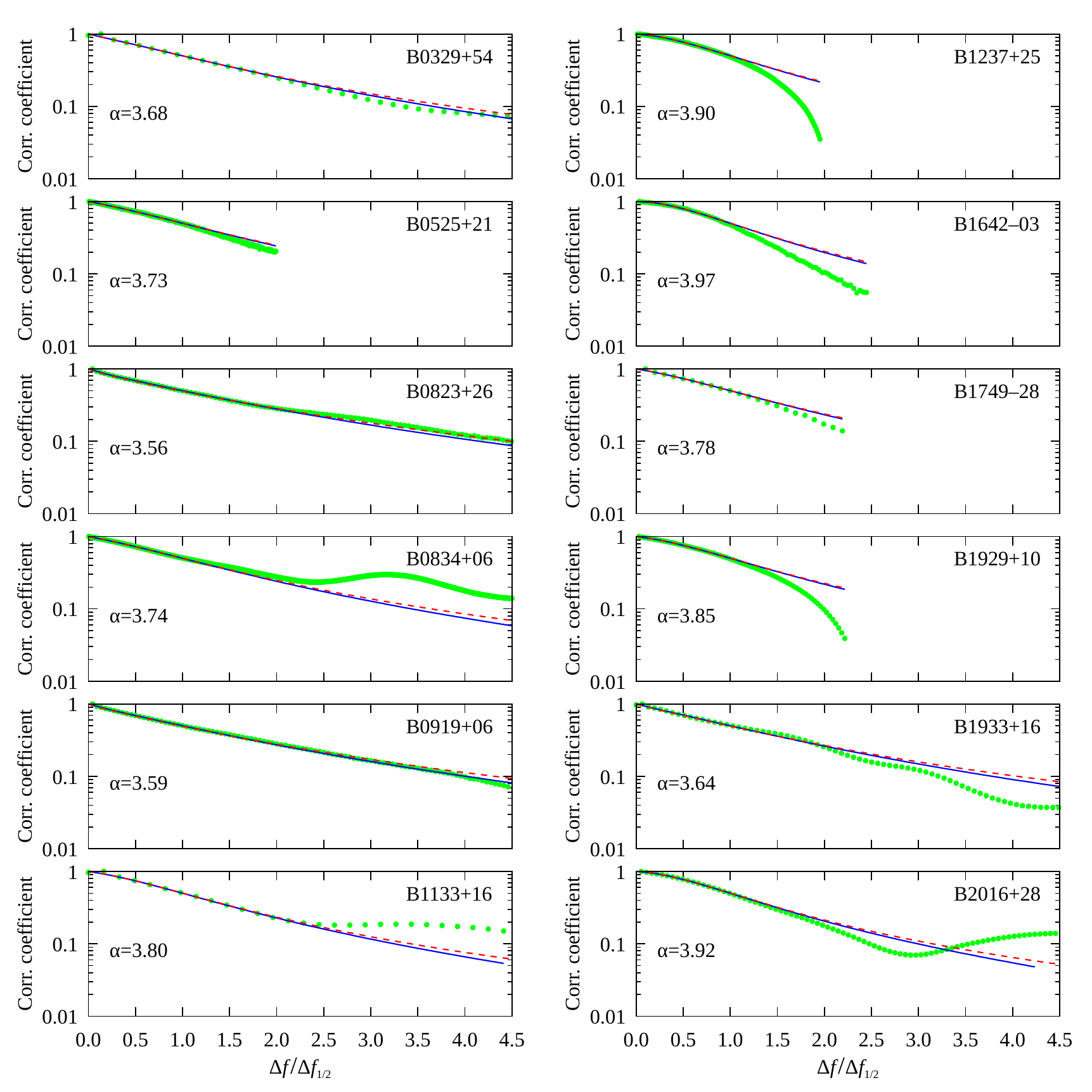}
\caption{Comparison of the $ACF(\Delta f)$ functions derived from
  our observations (green dots) with the best-fit
  theoretical predictions for the extended-medium model with
  power-law spectral index $\alpha$ as a fitting
  parameter (solid blue lines) and for the thin-screen model
  (dashed red lines) with the same spectral index for comparison. For better comparison all functions were
  normalized by their HWHM frequency lags, $\Delta f_{1/2}$. Only
  the inner part of the ACFs down to 45\% or up to 
  $\Delta f/\Delta f_{1/2}\sim 1.2$ for which a good fit could be
  obtained for all pulsars was used for the fit. The
  parameter, $\alpha$, which is essentially the same for our ACFs
  for both models for the fitting range is given for each
  pulsar. For some pulsars the
  width of the scintles in terms of $\Delta f_{1/2}$ relative to
  the receiver bandwidth of 16 MHz was relatively large so
  that only a shortened version of the functions was
  plotted.
\label{fig:fig5}
}
\end{figure*}

To search for any indications whether a thin-screen or an
extended-medium scattering model is preferred by our
observed $ACF(\Delta f)$ functions, we extend the functions to
frequency lags of 4.5$\times \Deltaf_{1/2}$ and plot the
observed ACFs from Figure~\ref{fig:fig3} together with the
predictions of the two scattering models in
Figure~\ref{fig:fig5}. For pulsars with broad scintles
relative to the bandwidth of 16 MHz (see, e.g., PSR
B0525+21, Figure~\ref{fig:4dynspectra} and
Table~\ref{tbl:results}), we plot the functions only for a
portion of the total range in frequency lags.

Slight differences in the models can be seen at $\Delta f\sim 2.5\times
\Deltaf_{1/2}$ that grow larger with further
frequency lags. For most pulsars the tail of the measured
ACFs deviates substantially from either of the models and,
as discussed in subsection \ref{subsec:freq_section}, can
likely be interpreted in most cases as being due to
deficiencies in the theoretical scattering models.

\begin{figure}[htb]
\includegraphics{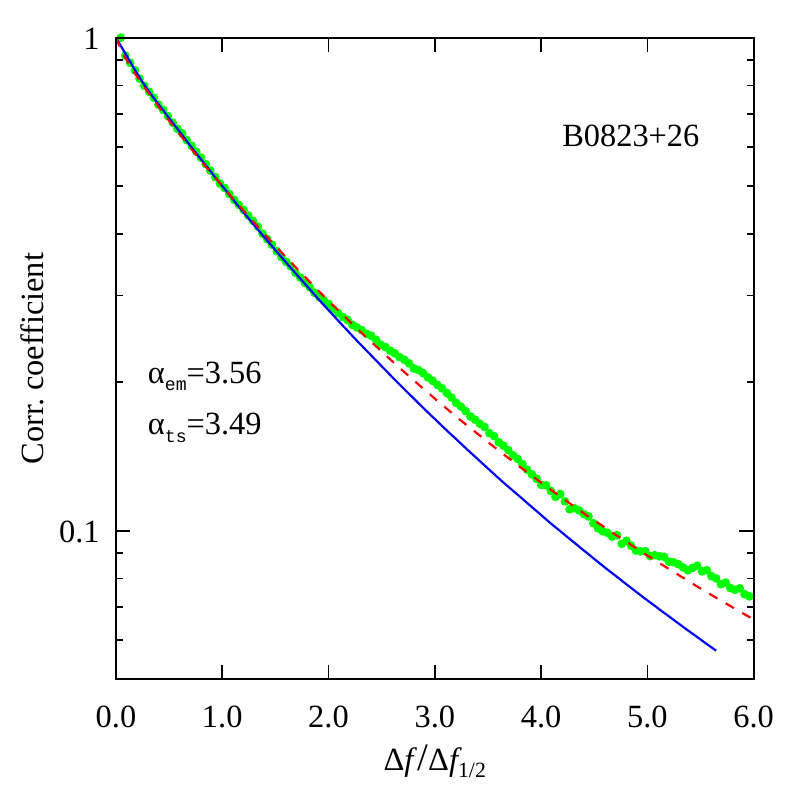}
\caption{The $ACF(\Delta f)$ function (thick green line) for PSR
  B0823+26 compared with the extended medium (solid blue
  curve) and thin-screen model (dashed red curve) but for
  frequency lags up to about $6\times\Deltaf/\Deltaf_{1/2}$. The
  extended medium model is drawn as in Figure
  ~\ref{fig:fig4} with $\alpha_\mathrm{em}=3.56$. The thin-screen model
  is plotted for $\alpha_\mathrm{ts}=3.49$ as a test, in an attempt to match the
  tail of the ACF without excessively changing the best-fit
  value of $\alpha=3.56$.
\label{fig:fig6}
}
\end{figure}

Three pulsars, B0329+54, B0823+26 and B0919+06, show fairly
good consistency in the observed ACFs with either, or both,
of the models for frequency lags at least up to
$4.5\times\Deltaf_{1/2}$. For PSRs B0329+54 and B0919+06,
deviations from either of the two models clearly grow for
even larger lags. PSR B0823+26, however, is a candidate that
warrants more scrutiny.  In Figure~\ref{fig:fig6} we show
the observed ACF and the two model curves but now up to
$\Delta f\sim6\times\Deltaf_{1/2}$. The extended-medium model
is drawn as in Figure~\ref{fig:fig5} with $\alpha_{em}=3.56$. The
thin-screen model curve lies for the same $\alpha$ value above
the extended-medium model curve and is therefore a candidate
for matching the tail of the ACF, but not that much that it
would be a good fit to the observed ACF. To investigate this
case further, we plot the thin-screen model for
$\alpha_{ts}=3.49$, 3.5 $\sigma$ below the best fit value in an
attempt to deviate not too much from the best fit but also
to match the tail of the ACF. However, even in this extreme
case, deviations much larger than for the inner part of the
ACF are still visible. We therefore think that also in this
case the deviations are due to effects discussed already for
the other pulsars.  There is no convincing case that either
of the two models is preferred.

An alternative approach to search for differences is to
analyze the Fourier transform of the ACFs,
$\mathcal{FT}[ACF(\Deltaf)]$ \citep[e.g.,][]{Armstrong1981,
  Wol1983} and the respective thin-screen and
extended-medium models as shown in Figure~\ref{fig:fig4}.
Differences between the models are somewhat more pronounced
than in Figure~\ref{fig:fig5}. However also in this case,
the observed functions $\mathcal{FT}[ACF(\Deltaf)]$ could
not conclusively discriminate between either of the models.

\subsection{Relation to the dynamic visibility function from space VLBI Radioastron observations}
The visibility function, $V(\tau,t)$, as a function of delay
and time in our space VLBI observations is obtained from the
inverse Fourier transform of the cross-spectrum between two
different radio telescopes.  In our case of observations of
dynamic spectra with single telescopes, the cross-spectrum
corresponds to the auto-spectrum at any particular time,
$t$, in a dynamic spectrum, $S(f,t)$.  A Gaussian has
been used to approximate the frequency section of
the ACF's of dynamic spectra \citep[e.g.][]{Cordes1986}.
Since for a Gaussian ACF, the underlying function and its
Fourier transform are also Gaussians, the auto-spectrum
would also be expected to be a Gaussian. 
Table~\ref{tbl:bells} lists the function $y$ as a Gaussian function, $G(f)$, together
with three other functions, $L(f)$, $E(f)$, $K(f)$, we consider further in our
analysis.  In addition we list for each function, $y$, its Fourier transform, $\mathcal{FT}(y)=\tilde y$, and its $ACF (y)$.

\begin{deluxetable*}{lLChC}[ht]
\tablecaption{\label{tbl:bells}Fourier transform and self-convolution of four parametric families of functions}
\tablehead{
 \colhead{Name} &
 \colhead{ $y$} 
& \colhead{$\tilde y$}  
& \nocolhead{$H[f]\cdot H[\tilde{f}]$}
&\colhead{$ACF(y)$}
}
\startdata
Gauss & 
  G(\sigma,f)=\displaystyle\frac{\exp{(-{{f^2}/({2\,\sigma^2}))}}}{\sqrt{2\pi}\,\sigma} 
& 
    \displaystyle\frac{G\left(1/(2\pi\sigma), t\right)}{\sqrt{2\pi}\, \sigma}
& \displaystyle\frac{1}{2\pi}
& G(\sqrt{2}\,\sigma,\Delta f) \vphantom{\Biggm|}\\
Lorentz & 
  L(\lambda, f)=\displaystyle\frac{\lambda}{\pi\left(\lambda^2+f^2\right)}
& 
   \displaystyle\frac{E\left(1/(2\pi\lambda), t\right)}{\pi\,\lambda} 
& \displaystyle\frac{1}{2\pi} 
& L(2\lambda,\Delta f)    \vphantom{\Bigm|} 
\\   
Laplace &  
  E(\epsilon,f)=\displaystyle\frac{\exp{(-|f|/\epsilon)}}{2\,\epsilon}
& 
  \displaystyle\frac{L\left(1/(2\pi\epsilon),\,t\right)^{\vphantom)}}{2\,\epsilon}
&  \displaystyle\frac{1}{2\pi}
& \displaystyle\frac{(\epsilon+|\Delta f|)E(\epsilon,\Delta f)}{2\,\epsilon}  \vphantom{\Bigm|}\\
 Bessel\quad\quad\quad &
K(\phi, f)=\displaystyle\frac{1}{\pi\phi}K_0\left(\frac{|f|}{\phi}\right)
& 
   \displaystyle\left(\frac{L\left(1/(2\pi\phi),\,t\right)}{2\,\phi}\right)^{1/2}
&  \displaystyle
&  E(\phi,\Delta f) \vphantom{\Biggm|}
\enddata
\tablecomments{Name of the function, $y$, definition of $y$, 
Fourier transform of $y$ with $\mathcal{FT}(y) = \tilde y$, 
The Fourier transform is defined
here as 
$\tilde{y}(p,t)= 
\int^{+\infty}_{-\infty}y(p,f)e^{-2\pi if t}\,\mathrm{d} f $, with $p$ as a parameter. 
All the
functions considered here are real
and normalized so that $\tilde y(0)=1$.  
Therefore, 
the autocorrelation function is identical with
self-convolution 
and $\widetilde{ACF}={\tilde y}^2$.
Because of the normalization chosen, the functions may be
identified with probability densities of random variables.
The functions named ``Lorentz'', ``Gauss'', and ``Laplace''
are related to the well-known statistical distributions
whose mathematical properties are described by
\cite{FellerII}.  For the last row, we used the term
``Bessel'' for the function since $K_0(z)$ is the modified 
Bessel function of the second kind of order
0 and argument z which rises logarithmically
to infinity at $f\to0$.  The scale parameters $\sigma$ and $\phi$ are
equal to rms deviations of the corresponding distributions,
and the parameters $\lambda$ and $\epsilon$ are half-widths of the Lorentz and
Laplace distributions at the level $1/2$ and $1/e$,
respectively.
}
\end{deluxetable*}

\begin{figure}[htb]
\includegraphics{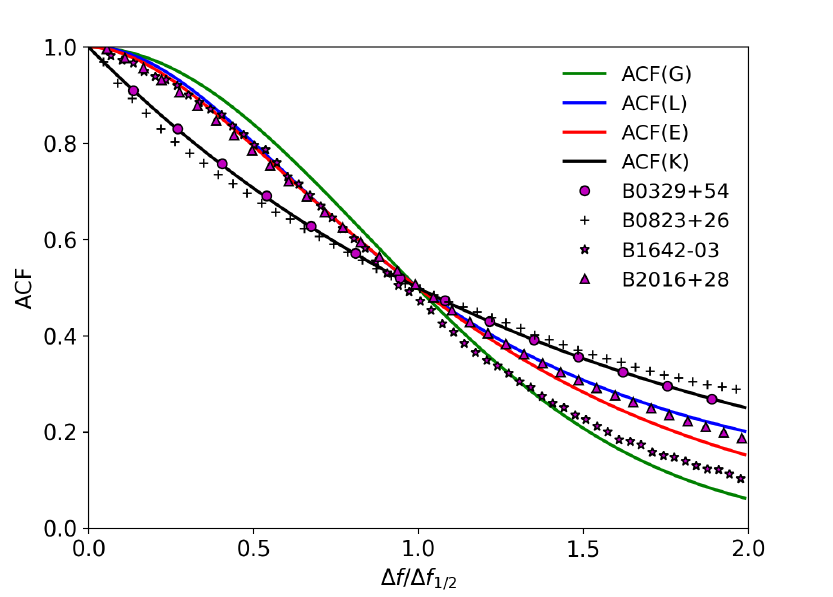}
\caption{The four $ACF(y)$ functions listed in
  ~Table~\ref{tbl:bells} and plotted as a function of
  frequency lag $\Delta f$. The functions are normalized in
  amplitude and with respect to their HWHM, $\Delta f_{1/2}$, for
  easy comparison. All functions are two-sided, only one
  side is plotted. In addition observed ACFs are plotted for
  four pulsars to display the range of ACF variations for
  comparison.
\label{fig:fig7}
}
\end{figure}

Space VLBI with Radioastron revealed for some pulsars that
the tail of the delay cross-section of $V(\tau,t)$ retained
significant magnitudes at baseline projections where the
pulsar scattering disk was completely resolved.  The delay
cross-section, $ACF(\Delta \tau,\Delta t=0)$ of $V(\tau,t)$, was found to be
well approximated by a Lorentz function \citep{pop2020}.
The Fourier transform of a Lorentz function is a two-sided
exponential and therefore the auto-spectrum from our
$S(f,t)$ would also be expected to be a two-sided
exponential (second row in Table~\ref{tbl:bells}).  The ACF
of such an exponential, $ACF(E)$, is a function given in the
third row in Table~\ref{tbl:bells}.

Independent of any prior knowledge from space VLBI
observation, an inspection of our observed ACF functions in
Figure ~\ref{fig:fig3} indicates that several of them have
up to $\sim0.5\times\Deltaf_{1/2}$ a concave profile, with a rounded
top where the second derivative is negative. Such a profile
is similar to the analytical functions, $ACF(G)$, $ACF (L)$
and $ACF(E)$ in Table ~\ref{tbl:bells}. Some others have in
contrast a convex inner profile where the second derivative
is positive. This profile is similar to $ACF(K)$.

In Figure~\ref{fig:fig7} we plot as examples the observed
ACFs of four pulsars that define the range in curvature for
all 12 pulsars, and compare them to the four analytical
$ACF(y)$ functions. PSR 1642-03 has the most concave inner
ACF of all pulsars. It is between that of the $ACF(G)$ and
$ACF(E)$ functions. For the inner part it resembles $ACF(E)$
and for the outer part of the plot in lies between $ACF(E)$
and $ACF(G)$. For PSR 2016+28, the ACF is closest to
$ACF(L)$ and $ACF(E)$ for the inner part and lies between
the two functions for the outer part. The least ambiguous
case is found for PSR 0329+54. The ACF is well matched by
$ACF(K)$. PSR B0823+26 has the most convex ACF even slightly
more so than $ACF(K)$. In Table ~\ref{tbl:results} we list
the rms values of the observed ACFs to the closest
analytical $ACF(y)$ functions and indicate which functions
they are. The residuals for all pulsars vary between rms
values of 0.0015 for B0329+54 and 0.037 for B1749-28. They
are all larger by a factor 2 to 9 than those for the
scattering models, however they are also from a fit over a
larger lag range and do not have a free parameter for
adjustment. In general the ACFs of all 12 pulsars are
approximately within the range of the four analytical
functions.

Does the ACF of any of these pulsars resemble a Gaussian? We
fit a function of the form, $f_g(\Delta f)=\exp(-|\Delta f|^{\delta}/b_f)$
to the ACF of PSR B1642-03 which is one of the two most
likely candidates for its ACF to have a Gaussian shape. We
obtained for the free exponent and its statistical error,
$\delta=1.79\pm0.03$, still significantly smaller than $\delta=2$ for
a Gaussian. However, in general it appears that for a few
pulsars with a concave shape of the inner ACFs there can be
a tendency in the shape of the ACF toward a Gaussian.

Listing the 12 pulsars according to the shape of their ACFs
it is clear that the most concave inner ACFs are related to
the highest values of $\alpha$ and the most convex to the lowest
value. PSR B0823+26 has the most convex ACF, even slightly
more so than $ACF(F)$, and the lowest value of $\alpha$ of 3.56.
At $\alpha=\sim3.75$ the ACFs transition from concave to convex.

\subsection{The average frequency profile of the scintles}

The average frequency profile of the scintles could
  be obtained analytically by finding the function, $y$ in
  Table~\ref{tbl:bells} if our observed ACFs were exactly
  represented by $ACF(y)$. However, only very few can be
  approximately associated with one particular $ACF(y)$, and
  then only for a limited range of frequency lags. Most are
  between neighboring analytical functions or are closer to
  one function in their inner parts and to another function
  at the tail. Therefore it is best to use the comparison of
  the observed ACFs with the analytical functions as a
  guide. The inferences as to the scintle profile can be
  only approximate.

\begin{figure}[htb]
\includegraphics{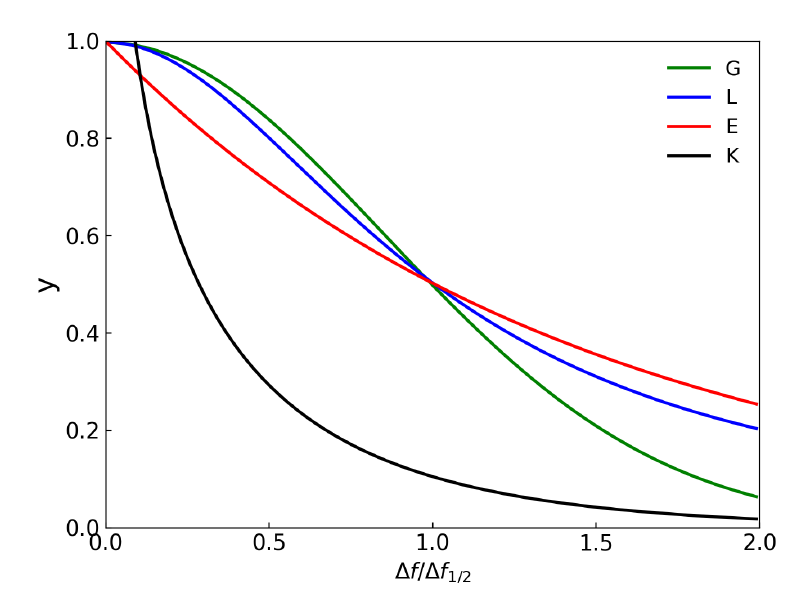}
\caption{The four functions, $y$ in Table~\ref{tbl:bells}
  plotted as a function of frequency, $f$. Three functions
  are normalized in amplitude and width to facilitate
  comparison. The fourth function, $K$, is plotted with the
  respective normalization parameter corresponding to the
  function, $E$. It goes to infinity at $f\to0$. As in Figure
  ~\ref{fig:fig7}, all functions are two-sided, only one
  side is plotted.
\label{fig:fig8}
}
\end{figure}

In Figure \ref{fig:fig8} we plot the functions,
  $y$, which are the undelying functions to $ACF(y)$. The
  functions $G$ and $L$ are concave up to $\sim$HWHM with a
  bell-shaped profile and the other two functions, $E$ and
  $K$ are convex with a profile of a cusp either with a
  moderate or an extreme peak. In the latter case, the
  function goes to infinity for $f\to0$.

Comparing the $ACF(y)$ with the $y$ functions in
  Figures \ref{fig:fig7}, \ref{fig:fig8} and
  Table~\ref{tbl:bells}, it is clear that the observed ACFs
  matched best by $ACF(G)$ and $ACF(L)$ functions are
  associated with bell-shaped scintle profiles. The $ACF(E)$
  function is almost identical to $ACF(L)$ for small lags
  and becomes clearly different for lags $\Delta
  f>\Deltaf_{1/2}$. However the function $y$ assumes the
  shape of a cusp. Five of our pulsars have clear
  similarities to $ACF(E)$ but also to any of the other two
  functions, $G$ and $L$. The scintle profile is probably
  described best as a hybrid between a bell-shaped curve and
  a cusp.

For one pulsar, B1929+10, the ACF is best described
  only by $ACF(E)$. None of the observed ACFs is best
  described only by the bell-shaped functions, G and L.  For
  three pulsars, B0525+21, B1133+16, and B1749+28, there is
  a transition from concave to convex ACF shapes. Their ACFs
  are hybrids with $ACF(K)$ as one element.  For five other
  pulsars their ACFs are clearly convex and best described
  by only $ACF(K)$. The scintle profile is less
  ambiguous. Although the function, $K$, goes to infinity
  for $f\to0$, because of deviations of the observed ACFs from
  $ACF(K)$, the scintle profile would probably be best
  described as a pronounced cusp with a sharp peak.

What is the relation of the wavenumber power-law spectral
index to the scintle profile? It appears that for pulsars
with high $\alpha$ values , that is steep wavenumber spectra, the
scintle could have a profile intermediate between a cusp and
a Gaussian or Lorentzian because of a tendency of the shape
of their ACFs toward these functions.  Below the transition
from a concave shape of the inner ACFs to a convex shape at
$\alpha\simeq 3.75$ the cusp becomes more clearly defined with a sharp
peak and its sides decrease even faster than exponentially
in the case of pulsars with the flattest wavenumber spectra.

\section{Discussion}
\label{sec:disc}

The electron density variations in terms of the wavenumber
power-law spectral index, $\alpha$, can be estimated by several
methods by, e.g., analyzing the shapes of ACFs of dynamic
spectra in frequency and time, the frequency dependence of
the decorrelation bandwidth, $\Delta f_{1/2}$, the scattering
time, $\tau_{sc}$, and the scattering angle, $\theta_{sc}$. We
estimated the index, $\alpha$, for 12 pulsars by analyzing the
frequency section of the time averaged ACF of dynamic
spectra and comparing it to extended-medium and thin-screen
scattering models. This is the larges sample we are aware of
that has been used for such analysis and has given us
important insight into characteristics of the turbulent
interstellar plasma and the associated frequency profile of
the scintles.

The inner part of the ACFs of all 12 pulsars down to at
least 45\% of the maximum or $1.2\times\Delta f_{1/2}$ 
is for most pulsars well fit by the predictions of
extended-medium 
and thin-screen
 scattering theories.  The
power-law spectral indices range from $\alpha=3.56$ to 3.97 with
errors $\leq0.05$ and a mean with standard deviation of
3.76$\pm$0.13 which is close to the index of 3.67 for the
Kolmogorov spectrum.  
The largest deviations from either of
the models were found for PSR B1642-03 with an error for $\alpha$
of 0.05.

All pulsars show deviations of their ACFs from either of the
models at frequency lags starting at $\sim1.2$ to $3\times\Delta
f_{1/2}$.  These deviations are larger than any found for
the inner part of the ACFs.  For some of them the ACFs turn
to a level above the models which could possibly be due to
technical reasons.  However for most of them the ACFs turn
to a level clearly below the models which likely has
physical reasons.  A more compact ACF is equivalent to a
broader $\mathcal{FT}(ACF)$ (compare Figures \ref{fig:fig4}
and \ref{fig:fig6}).  Longer delays correspond to more
compact inhomogeneities which may indicate deficiencies in
either of the two scattering models.

The most direct comparison of our measurements of $\alpha$ are
those obtained from $\mathcal{FT}[ACF(\Deltaf)]$. While the
estimates of \citet{Armstrong1981} have rather large allowed
ranges, the estimates of \citet{Wol1983} (no errors given),
are equal within our errors of 0.1.  Although only three
pulsars could be compared, this result is remarkable, since
the observations were made some thirty years apart and therefore give
hints as to the stability of the power-law spectrum for the direction of these pulsars. 

Of particular interest is a comparison of an estimate of $\alpha$
by \citet{Smirnova21} who used for 10 of our pulsars the
same dynamic spectra but focused on the time section of the
frequency averaged ACFs and analyzed the structure
function. While for five pulsars their values are only
somewhat larger or smaller within a difference of $\leq$ 0.10,
for the other pulsars their values are all smaller by as
much as 0.51 in the case of B1237+25.  Such large
discrepancies are remarkable given that the estimates were
obtained on the basis of the same spectra but separately for
their frequency and their time characteristics.

The estimates of $\alpha$ from the frequency dependence of $\Delta
f_{1/2}$ as given in our example in Table \ref{tbl:compare}
were largely consistent with our values, albeit within the
authors' \citep{cwb1985} larger errors.

Independent information on $\alpha$ comes from measurements of
the exponential broadening of pulses through multiple path
propagation in the ionized ISM. The parametrization is
generally given by the broadening time scale or scattering
time, $\tau_{sc}$.  This parameter in turn is related to
$\alpha$. For a power-law wavenumber spectrum, $\tau_{sc}\propto f^{-\beta}$
and for $\alpha < 4$, $\beta = 2\alpha/(\alpha-2)$. That requires
$\beta<4$. \citet{Loehmer+2004} found for seven pulsars
$3.3\leq\alpha\leq4.0$ and for PSR B1933+16 in our list, $\beta=3.4\pm0.2$,
too small for $\alpha < 4$. \citet{Bhat+2004} found for eight out
of 15 pulsars $3.1\pm0.7\leq\alpha\leq4\pm1$ but for the others also
values of $\beta$ too small for $\alpha <4$. An even larger portion
of pulsars outside the $\alpha$ range was found by
\citet{Geyer2017} with $1.5 \leq\beta \leq 4.0$ for 13 pulsars and
only two of them with $\alpha<4$ within the errors. The largest
sample of pulsars with multifrequency observations was
obtained by \citet{Levand2015}. From 48 pulsars only 16 were
given with values of $\alpha$ within their errors $<4$ including
the only one from our list, PSR B1749-28 with
$\alpha=4.24\pm0.18$, different from our value of 3.78$\pm$0.02
by 3 times their larger uncertainty.  All the others
including PSRs B1642-03 and B1933+16 from our list, had
correspondingly smaller values of $\beta$, with an average of
3.89.

The consistent result from pulse broadening observations is
that a large fraction of pulsars have $\beta$ values too
small to be consistent with $\alpha<4$. Several effects
could be considered that would lead to an increase or
changes of $\beta$. There is an indication that pulsars with
$DM<500$ pc cm$^{-3}$ have with $\beta=3.95$ a larger
average than those with $DM>500$ pc cm$^{-3}$ with
$\beta=3.49$ \citep{Levand2015}. All of our pulsars have in
comparison very small dispersion measures (see, Table
\ref{tbl:results}). \citet{Geyer2017} find that their small
values of $\beta$ may be indications of anisotropic
scattering since that assumption would lead to an increase
of the values. Anisotropic scattering mechanisms have also
been considered by \citet{Stin2001, TuntsovBW2013}. Other
effects discussed include finite or truncated scattering
screen \citep{CordLaz2001} and internal cutoff scale effects
\citep{RickBar2009}. Also, analysis of the structure
function of several pulsars, based on multifrequency
observations, showed that the spectrum of interstellar
plasma in the direction of some pulsars follows a piece-wise
power law \citep{Shishetal2003, Smirnova2006}. Our
measurements of 12 pulsars with a different analysis scheme
adds to the discussion of the wavenumber spectrum and the
electron density variations of the plasma turbulence of the
ISM. In general it appears that more complex theoretical
models are needed to describe the observed data or possibly
that there are deviations from the wavenumber power-law spectrum.

In our search for the average frequency profile of the
scintles in terms of analytical functions independent of
scattering models we were guided by our earlier results from
VLBI observations of pulsars where we found that the delay
section of the visibility function of some pulsars could be
well fit by Lorentzians. For half of our pulsars we indeed
found that the ACF's could be best fit by the corresponding
$ACF(E)$ function or by a hybrid of functions with $ACF(E)$
being part of it.  However, for five others the more sharply
pointed two-sided exponential, $ACF(K)$ was warranted in
addition to three more where that function was part of a
hybrid. It is interesting to note that while the set of
power-law spectral indices from $\alpha=3.97$ to 3.56 appears to
be a uniform distribution, the functions describing the ACFs
and the shape of the scintles are at least formally quite
different. For the steep spectra with $\alpha \gtrsim 3.75$, the inner
part of the ACFs become increasingly concave while for the
flatter spectra with $\alpha\lesssim3.75$ they become convex. In other
words, the value of $\alpha$ determines the average frequency
profile of the scintles. Steep wavenumber spectra with $\alpha <
4$ correspond to scintles with a somewhat rounded cusp. With
smaller $\alpha$ values the peak of the cusp becomes more
pronounced. For $\alpha\lesssim3.75$ and further flattening, the cusp
and its peak sharpen further and decay faster than an
exponential, approaching at least nominally the modified
Bessel function of the second kind of order zero.

Is there any correlation between the shape of the scintles
and any of the pulsar characteristics listed in
Table~\ref{tab:param0} such as the dispersion measure, the
distance to the pulsar and the galactic coordinates? We
searched for such a correlation but no correlation is
apparent.

\section{Conclusions}   \label{conc}
\begin{enumerate}

\item We analyzed the dynamic spectra of 
nine pulsars at a
  center frequency of 324~MHz and three pulsars at 1676~MHz
  and computed the frequency sections of the two dimensional
  autocorrelation functions.

\item For each pulsar the inner part of the function down to
  at least 45\% of the maximum is well fit by the prediction
  of a 
thin-screen or
 extended-medium scattering model. The power
  law wavenumber spectral indices of the interstellar plasma
  turbulence, $\alpha$, are all within a range of 3.56 and 3.97
  with uncertainties $\leq0.05$.
  
\item 
The mean of the spectral indices with standard
  deviation is $\langle\alpha\rangle = 3.76 \pm0.13$ which is close to the
  Kolmogorov index of 3.67.
  
\item 
The Fourier transforms of the model functions fit to
  those of the ACFs for the full width gives similar values
  for alpha, although with larger uncertainties.

\item Beyond the inner part of the function, clear misfits
  can be seen 
for all of our pulsars that are larger than
  any possible deviations seen in the inner part of the
  ACFs,
 indicating scattering characteristics more complex
  than described in the models, 
or indicating that there are
  deviations from the power law of the interstellar plasma
  turbulence.
  
\item 
Comparison of extended-medium and thin-screen models
  with observed ACFs and the respective $\mathcal{FT}(ACF)$s
  gives no clear evidence that either of the models is
  preferred.

\item 
The observed ACFs have a concave inner part down to about half of the maximum for high $\alpha$ values that becomes less concave with flattening spectra and turns convex for $\alpha\lesssim3.75$.

\item 
  For six pulsars, with $3.75\lesssim\alpha<4.0$, the
  function $ACF(E)$ alone or as a member of a hybrid fits
  the observed ACFs moderately well down to 20\% of
  the maximum. This function is expected for pulsars for
  which the Lorentzian provides a fairly good fit to the
  visibility function from VLBI.

\item 
For the pulsars, with $3.56\leq\alpha\lesssim3.75$ a
  function, like the two sided exponential, $ACF(K)$, is more
  warranted for the fit to the ACFs.
  
\item 
 A Gaussian was not an appropriate fit for any of
  the ACFs of the 12 pulsars. The pulsars who came closest
  are B1237+25 and B1642-03 with ACFs described best by a
  hybrid between $ACF(G)$ and $ACF(E)$.

\item From the functional fit to the observed ACFs we found
  that the average frequency profile of a scintle is
  for steep wavenumber spectra characterized by a
    hybrid between a cusp and a function like a Gaussian or
    Lorentzian.  With increasing flattening of the spectrum
    with $\alpha\lesssim3.75$, the cusp and its peak becomes more
    pronounced and decays faster than an exponential.

\end{enumerate}

\acknowledgments We thank an unknown referee for valuable
comments and suggestions. The Radioastron project is led by
the Astro Space Center of the Lebedev Physical Institute of
the Russian Academy of Sciences and the Lavochkin Scientific
and Production Association under a contract with the Russian
Federal Space Agency, in collaboration with partner
organizations in Russia and other countries. This paper was
supported in part by the Russian Academy of Science Program
KP19-270 ``The study of the Universe origin and evolution
using the methods of earth-based observations and space
research''. NB was supported by the National Sciences and
Engineering Research Council of Canada.
\facilities{SRT, GBT, Aresibo, 
WSRT, Pushchino 22-m radio telescope (used as Tracking Station)}.
\software{CFITSIO}

\bibliographystyle{aasjournal}
\bibliography{SE-PSR}
\end{document}